\documentclass[twocolumn,10pt]{asme2e}

\usepackage{ifthen}

\usepackage{hyperref}
\usepackage{amsmath,mathrsfs,amssymb,amsfonts}
\usepackage{bm}

\usepackage{subfigure}

\usepackage{graphicx}
\usepackage{epsfig}
\usepackage{psfrag}

\usepackage{pstricks}
\usepackage{pstricks-add}
\usepackage{pst-math,pst-xkey,pst-plot}

\usepackage[open,atend]{bookmark}
\bookmarksetup{color=black}
\bookmarksetup{openlevel=1}

\usepackage{tabularx}
\graphicspath{{figures/}}
\def \figpath {./figures}

\confshortname{IDETC/CIE 2009}
\conffullname{ASME 2009 International Design Engineering Technical Conferences \&\\
              Computers and Information in Engineering Conference}
\confdate{August 30 - September 2}
\confyear{2009}
\confcity{San Diego}
\confcountry{USA}

\papernum{DETC2009-87603}

\title{2D MULTI-OBJECTIVE PLACEMENT ALGORITHM FOR FREE-FORM COMPONENTS}

\author{Guillaume Jacquenot\thanks{Address all correspondence to this author.}\ $^{(1,2)}$, Fouad Bennis$^{(1)}$, Jean-Jacques Maisonneuve$^{(2)}$, Philippe Wenger$^{(1)}$
    \affiliation{
	$^{(1)}$ \'{E}cole Centrale de Nantes, IRCCyN, UMR CNRS 6597, Nantes, France\\
	$^{(2)}$ SIREHNA, Nantes, France\\
    Email: guillaume.jacquenot@gmail.com
    }	
}

\hypersetup{
   pdftitle    = {2D MULTI-OBJECTIVE PLACEMENT ALGORITHM FOR FREE-FORM COMPONENTS},
   pdfsubject  = {ASME 2009},
   pdfauthor   = {JACQUENOT Guillaume},
   pdfkeywords = {Placement problems, Cutting and Packing problems, Layout problems, Multi-objective optimization},
   }

\let\captionOld\caption
\renewcommand{\caption}[1]{\captionOld{\MakeUppercase{#1}}}

\newcommand{\etal}{\emph{et~al.\ }}
\newcommand{\ie}{\emph{i.e.\ }}
\newcommand{\eg}{\emph{e.g.\ }}

\renewcommand{\theenumi}{$\bullet$}

\begin{document}

\bookmarksetup{bold}
\pdfbookmark[0]{2D MULTI-OBJECTIVE PLACEMENT ALGORITHM FOR FREE-FORM COMPONENTS}{TitleDocument}
\bookmarksetup{bold=false}

\maketitle

\begin{abstract} 
{\it This article presents a generic method to solve 2D multi-objective placement problem for free-form components.
The proposed method is a relaxed placement technique combined with an hybrid algorithm based on a genetic algorithm and a separation algorithm.
The genetic algorithm is used as a global optimizer and is in charge of efficiently exploring the search space.
The separation algorithm is used to legalize solutions proposed by the global optimizer, so that placement constraints are satisfied.
A test case illustrates the application of the proposed method. Extensions for solving the 3D problem are given at the end of the article.
}
\end{abstract}

\section{Nomenclature}
\begin{center}
\begin{tabularx}{\linewidth}{lX}
	$m$ 	&Number of components\\
	$\bm v$&Vector of positioning variables \\
	$\bm v_i$&Vector of variables containing position variables of component $i$. $\bm v = \left(\bm v_1,\ldots,\bm v_m\right)$ \\
	$O_i$&Component $i$ \\
	$n_i$&Number of circles of component $i$ \\
	$A_i$&Area of component $i$ \\
	$\rho_i$&Density of component $i$ \\
	$S_{ij}$&$j^\textrm{th}$ circle of the $i^\textrm{th}$ component \\
	$\bm c_{ij}$&Vector of coordinates of the $j^\textrm{th}$ circle of the $i^\textrm{th}$ component \\
	$f_{ijkl}^{{\textrm{pen}}}$&Penetration penalty between the $j^\textrm{th}$ circle of the $i^\textrm{th}$ component and the $l^\textrm{th}$ circle of the $k^\textrm{th}$ component \\
\end{tabularx}
\end{center}
\begin{center}
\begin{tabularx}{\linewidth}{lX}
	$f_{ij}^{{\textrm{pro}}}$&Protrusion penalty of the $j^\textrm{th}$ circle of the $i^\textrm{th}$ component \\
	$C$&Container / Enclosure \\
	$\textrm{cl}(S)$&Closure of $S$ \\
\end{tabularx}
\end{center}

\section{Introduction}
The problem of placing a set of components inside an enclosure is known as the placement problem.
A placement problem asks to find all placement variables of
all components so that objectives are minimized and constraints are
satisfied. Solving a placement problem consists in finding one or several solution
that minimizes the objectives and respects a set of constraints.
Among all constraints, each placement problem presents
non-overlap and non-protrusion constraints. These constraints
respectively express the fact that components should not collide with each
other and that each component must lay inside the boundaries of the container.
These problems are non-linear and most of the time $\mathscr{N\!\! P}$-complete,
meaning that solutions associated with the corresponding decision problem can be checked with a polynomial algorithm.

Placement problems gather Cutting \& Packing (C\&P) problems and layout problems.
In a C\&P problem, components are only geometrically related to each other,
whereas in a layout problem, components are geometrically and functionnally related to each other \cite{Alad_07_obj1}.
The underlying objective of each C\&P problem is a compaction objective: either a maximum number of components has to be
placed in the container, or a minimum number of containers has to be used to place all components.
This characteristics, that may also be found in a layout problem, can influence the choice of modelling.

W\"ascher \etal \cite{Wasc_06_imp} recently proposed a typology of C\&P problems and
improved the one proposed by Dyckhoff \cite{Dyck_90_typ}.
Among all C\&P problems, knapsack problems \cite{Kell_04_kna},
bin packing problems \cite{Carl_07_new},
nesting problems \cite{Egeb_07_fas,Gome_06_sol}
container loading problems \cite{Pisi_02_heu} may be the most representative.
Layout problems also gather a wide panel of problems, including
Facility Layout (FL) problems \cite{Drir_07_fac},
Very Large Scale Integration (VLSI) problems \cite{Egeb_03_pla} and engineering problems \cite{Zhan_08_lay}.

Cagan \etal \cite{Caga_02_sur} proposed a survey on 3D layout problem, in which the
different aspects of layout problems are discussed.
Grignon \etal \cite{Grig_04_gab} proposed a multi-objective genetic algorithm framework to solve 3D layout problems, where individuals are clouds of solutions.
Aladahalli \etal \cite{Alad_07_obj1,Alad_07_obj2} developed a pattern search
algorithm to solve 3D layout problems, where objectives and constraints are
aggregated in a single merit function.
Yi \etal \cite{YiFa_07_veh} used a genetic algorithm (NSGA-II) to find the positions of a set
of components of a trunk. CAD is used to model the problem and perform geometric computations.
Non-overlap and non-protrusion constraints act as a penality when computing the rank of each solution.
Tiwari \etal \cite{Tiwa_08_fas} proposed a 3D bottom-left-back strategy to solve a 3D C\&P problem.
The voxelization of components allows to pack complex geometry components inside the container.
A steady-state genetic algorithm is used to generate the packing order and find the correct orientation of each component.
Placement constraints are automatically satisfied with the voxel-based representation.
Zhang \etal \cite{Zhan_08_lay} used a combination of soft computing techniques (GA/PSO) to place
a set of components in a satellite module under behavioral constraints. Components are modeled by either parallelepipeds or cylinders,
which allow using analytical functions to evaluate overlap and protrusion constraints.
Dong \etal \cite{Dong_06_veh} proposed a shape-morphing method to design components and solve a layout problem at the same time.

The objective of this article is to introduce a new placement method adapted to multi-objective problems.
The first section gives a general presentation of different placement methods, the second section concerns the proposed method.
An example is discussed in the third section, finally the different elements needed to develop the 3D placement method are presented.

\section{Placement methods}
Placement problems have generated a large amount of literature, however all placement techniques proposed can be classified in two categories:
legal placement method and relaxed placement method.
A legal placement method is a method that ensures that all placement constraints are satisfied while building the solution.
This is typically the case when using an encoding scheme (\eg bottom-left heuristic).
Tiwari \etal \cite{Tiwa_06_sur} proposed a survey of various encoding schemes, which can be used to propose different placements.
These placement techniques are perfectly suited for problems presenting compaction objectives,
such as the minimization of the wire-length of components inside an electronic module.
Relaxed placement methods allow the non-respect of placement constraints during the elaboration of the solution. These methods
are mainly used when dealing with complex geometries and can easily model any placement constraints.

When using a legal placement method, the decision variables of the problem are usually permutations,
which indicate the order of introduction of the components inside the container.
Other more evolved encoding schemes have been proposed such as the sequence pair encoding,
in which permutations introduce topological relations between components.
The decision variables of relaxed placement techniques are directly
the positioning variables of the components.

\begin{figure}[!htbp]
\centering
	\psfragscanon
	\psfrag{f1}[c][c]{$\phi$}
	\psfrag{a}[c][c]{(a)}
	\psfrag{b}[c][c]{(b)}
	\psfrag{f1}[c][c]{$f_1$}
	\psfrag{f2}[c][c]{$f_2$}
	\psfrag{x1}[c][c]{$x_1$}
	\psfrag{x1min}[c][c]{$x_{1\min}$}
	\psfrag{x1max}[c][c]{$x_{1\max}$}
	\psfrag{x2}[c][c]{$x_2$}
	\psfrag{x2min}[r][r]{$x_{2\min}$}
	\psfrag{x2max}[r][r]{$x_{2\max}$}
	\includegraphics[width=0.9\linewidth]{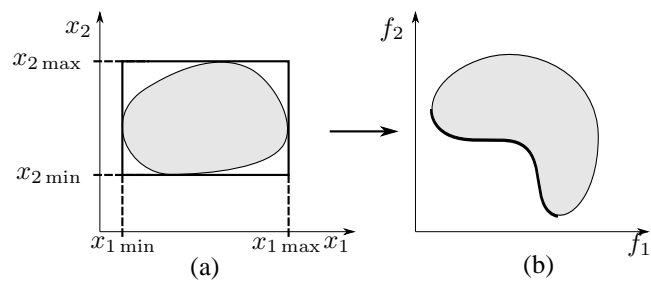}
    \caption{Illustration of the variable search space and objective search space for a bi-objective minimization problem.
             Thick dark line on subfigure (b) represents Pareto front of the problem, \ie the set of non-dominated points.}
\label{fig:Variables_2_Front_pareto}
\end{figure}

Placement problems are generally multi-objective problems.
However, most of the problems are treated as single-objective unconstrained problems,
where objectives and
contraints are aggregated in a single merit function.
Such a formulation has several shortcomings, first of all only the convex
part of the Pareto front can be identified. Second, the set of weights
used to aggregate functions can be difficult to identify, and may lead on
strong assumptions on the solutions obtained.
A multi-objective formulation is therefore preferred:
multi-objective algorithms will search for the set of non-dominated points
in the objective space given by efficient solutions,
as shown on figure \ref{fig:Variables_2_Front_pareto}.

\section{The proposed method}
The objective of the proposed method is to be generic, thus suited to a wide set of placement contraints and objectives.
A relaxed placement technique is therefore chosen.
The variables of the problem are directly the positioning variables of the components.
These variables can be of different types.
Translation variables are modeled as continuous variables and orientation variables
can be continuous or discrete depending on the designer's needs.

The difficulty of such a modelling will be to propose feasible solutions, \ie solutions that satisfy placement constraints.
The proposed method is based on the use of a separation algorithm, which objective is to legalize unfeasible solutions.
A global optimizer is used to generate different promising solutions.
The different elements needed for the resolution of the problem are presented in the following subsections.

\subsection{Geometry handling}
When using a relaxed placement method, collision or overlap detection may represent most of the computational effort.
The objective is therefore to use a correct geometric representation, that will be suited to our needs.

Several techniques can be used to detect a collision between two components.
The first one consists in computing the overlap area of each pair of polygons. However, this technique is too costly.
No-fit polygon (\emph{NFP}) and inner-fit polygon (\emph{IFP}) can also be used \cite{Benn_08_com}.
NFP and IFP present the advantage of allowing collision detection with a simple test of a point being in a polygonal region.
These features depend on the orientation of the
components, meaning that these elements have to be recomputed for each orientation.
When components are convex, this can be done in linear time with respect to the number of vertices of polygons.
Otherwise, a convex decomposition is needed.
However, this approach can hardly be adapted to 3D problems even in the case of polytopes.

\begin{figure}
	\centering
	\subfigure[Original shape]{
		\includegraphics[width=0.27\linewidth]{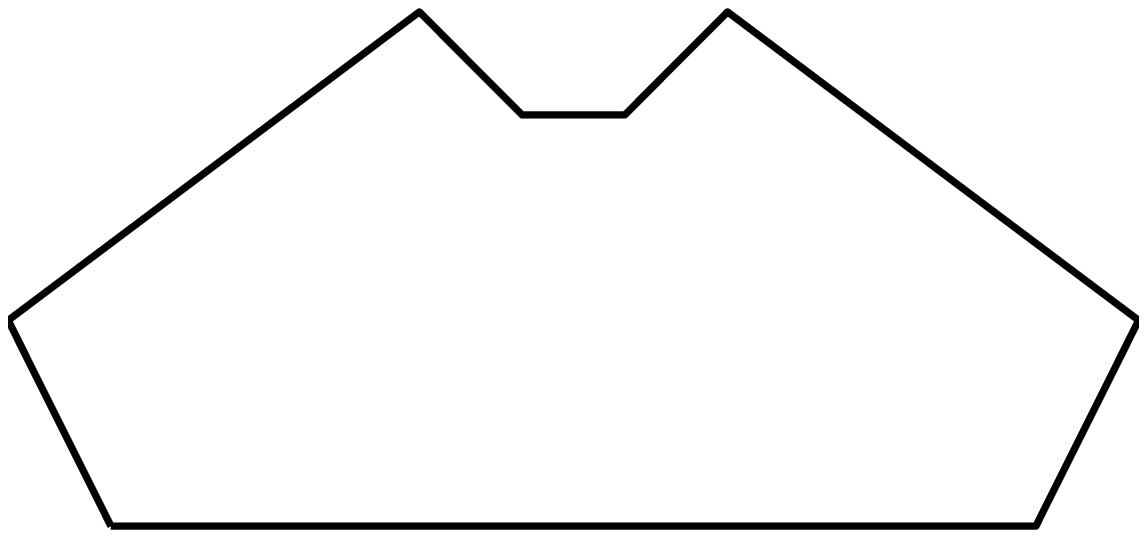} 
		\label{fig:Skeleton_a}
	}
	\quad
	\subfigure[Medial axis]{
		\includegraphics[width=0.27\linewidth]{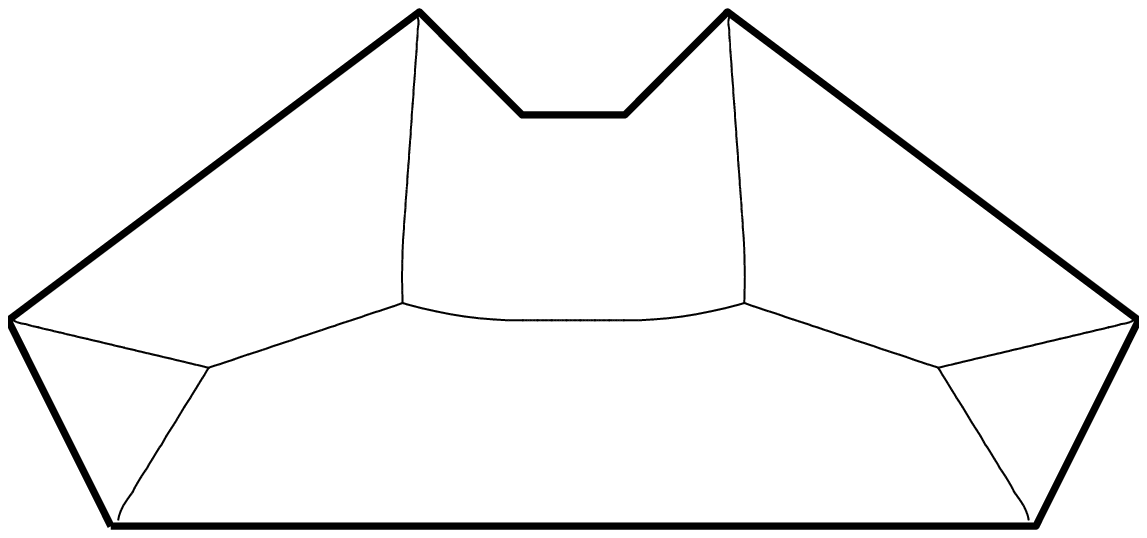} 
		\label{fig:Skeleton_b}
	}
	\quad
	\subfigure[Circle placement]{
		\includegraphics[width=0.27\linewidth]{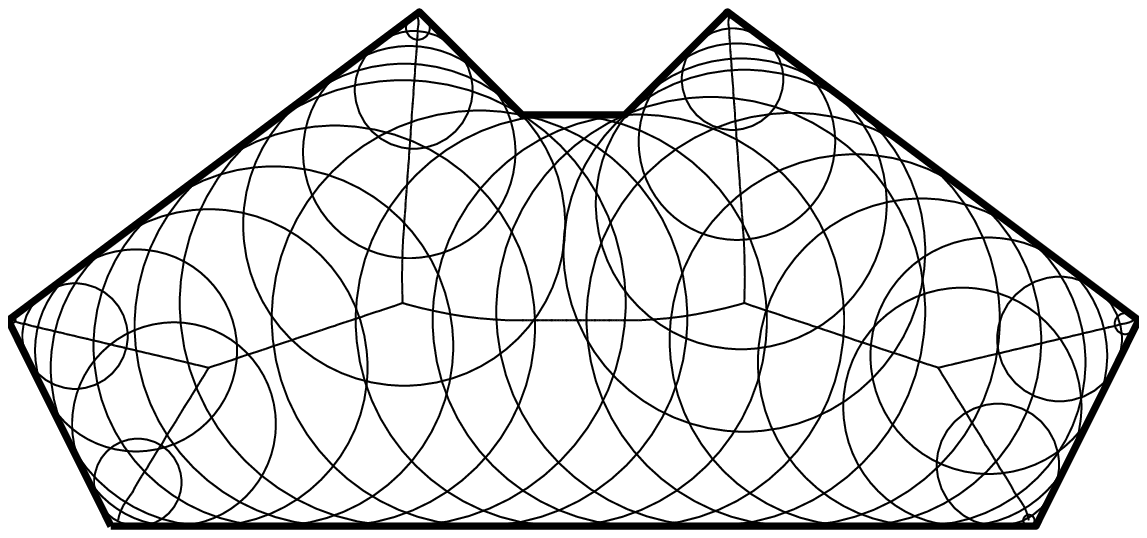} 
		\label{fig:Skeleton_c}
	}
    \caption{Different steps of the conversion of a polygon to a set of circles.}
    \label{fig:Skeleton}
\end{figure}

In case one is interested in the free rotation of components, components can be transformed into sets of circles.
The transformation proposed is based on the use of medial axis, which corresponds to the centers of maximum inscribed circles.
Figure \ref{fig:Skeleton} illustrates how components are converted to sets of circles.
First, the medial axis of the polygon is computed.
Second, a set of circles which centers are located on the medial axis are inserted.
Circles are inserted such that the distance between two centers is lower than a distance chosen by the designer.
We choose this representation for which collision detections between components can be performed with simple distance computations, and for which a separation algorithm has been proposed.

\subsection{Separation algorithm}

Different separation algorithms have been proposed, \cite{Imam_07_ite,Imam_07_mul,Imam_08_rem,Imam_08_des} however the key idea is always the same:
given a configuration that does not satisfy placement contraints, the objective of the separation algorithm is to minimize the non-respect of overlap and protrusion constraints.
These constraints are gathered in a penalty function $F$ characterizing the violation of the placement constraints.

The separation algorithm and the formalism used here are the ones proposed by Imamichi \etal \cite{Imam_08_des}.
Let's consider a collection of $m$ components $\mathscr O=\left\{O_1,\ldots,O_m\right\}$. Each component $O_i$ consists of $n_i$ circles $\left\{S_{i1},S_{i2}\ldots,S_{in_i}\right\}$.
Let $\bm c_{ij}$ be the vector that represents the center of circle $S_{ij}$, $r_{ij}$ be the radius of $S_{ij}$ $\left(i=1,\ldots,m; j=1,\ldots,n_i\right)$.
Circles are denoted with the capital letter $S$, because the separation algorithm can also be applied in 3D with Spheres.

The separation problem is an unconstrained minimization problem defined by
\begin{equation}
(\textrm{Sep})\left\{ {\begin{array}{l}
   {\displaystyle{\min\ F\left( \bm v \right) = \omega_\textrm{pen}F_\textrm{pen}\left( \bm v \right)+\omega_\textrm{pro}F_\textrm{pro}\left( \bm v \right) }}  \\
   {\textrm{s.t.} \quad \begin{array}{|l}
   \displaystyle{\bm v = \left(\bm v_1,\ldots,\bm v_m\right)} \\ 
\end{array}}\\
\end{array}} \right.
\label{eq:Min_PD_ILS_RIGID}
\end{equation}
where vector $\bm v$ represents the placement variables, which represent degrees of freedom (\emph{dof}) of the components.
$\omega_{\textrm{pen}}$ and $\omega_{\textrm{pro}}$ are positive parameters representing the weights of the penetration and protrusion function.
These values are respectively set to $1/3$ and $2/3$. The total penetration and protrusion penalty functions can be mathematically written as follows:
\begin{equation}
	\begin{split}
		F_{{\textrm{pen}}}\left( \bm v \right)  &= \sum\nolimits_{1 \leqslant i < k \leqslant m} {\sum\nolimits_{j = 1}^{n_i } {\sum\nolimits_{l = 1}^{n_k } {f_{ijkl}^{{\textrm{pen}}} \left(\bm v \right)} } } \\
		F_{{\textrm{pro}}}\left( \bm v \right)  &= \sum\nolimits_{i = 1}^m {\sum\nolimits_{j = 1}^{n_i } {f_{ij}^{{\textrm{pro}}} \left(\bm v \right)} }
	\end{split}
\end{equation}
where $f_{ijkl}^{{\textrm{pen}}}$ and $f_{ij}^{{\textrm{pro}}}$ represent the local penalties of the penetration and protrusion, defined by the penetration depth.
The penetration depth $\delta$ \cite{Agar_00_pen} of two components $O_1$ and $O_2$ is defined as the minimum translation distance of component $O_2$ so that $O_1$ and $O_2$ do not overlap.
These penalties write
\begin{equation}
	\begin{split}
		f_{ijkl}^{{\textrm{pen}}} \left(\bm v \right) 	&= \left( {\delta \left( {S_{ij} \left( {\bm v_i } \right),S_{kl} \left( {\bm v_k } \right)} \right)} \right)^2  \\
		f_{ij}^{{\textrm{pro}}} \left(\bm v \right) 	&= \left( {\delta \left( {S_{ij} \left( {\bm v_i } \right),{\textrm{cl}}\left( {\overline C } \right)} \right)} \right)^2
	\end{split}
\end{equation}
The penetration penalty of two circles $S_{ij}$ and $S_{kl}$ can be finally written as:
\begin{equation}
	f_{ijkl}^{{\textrm{pen}}} \left(\bm v \right) = \left( {\max \left\{ {r_{ij}  + r_{kl}  - \left\| {\bm c_{ij} \left( {\bm v_i } \right) - \bm c_{kl} \left( {\bm v_k } \right)} \right\|,0} \right\}} \right)^2
\end{equation}
Regarding non-protrusion constraints, inward offsets are used to evaluate if circles belong to the container. These constraints can be checked with a simple test of a point being inside a polygon. If the center of a circle is not contained inside the inward offset polygon, the minimum distance between the center and this polygon is used to build the constraint function and its gradient.

The function $F$ is a piecewise continuous function, therefore its gradient can be used to find its minimum. The gradient of function $F$ is
\begin{equation}
	\nabla F\left( \bm v \right) = \omega_\textrm{pen} \nabla F_\textrm{pen}\left( \bm v \right)+\omega_\textrm{pro} \nabla F_\textrm{pro}\left( \bm v \right)
\end{equation}
The gradient $\nabla F\left(\bm v \right)$ is computed by evaluating all components of gradients $\nabla F_\textrm{pen}$ and $\nabla F_\textrm{pro}$ as follows:
\begin{equation}
	\begin{split}
	\nabla_i F_\textrm{pen}\left(\bm v \right) &= \frac{{\partial F_{\textrm{pen}}\left(\bm v \right)}}{{\partial \bm v_i }} = \sum\limits_{j = 1}^{n_i } {\sum\limits_{k = \left\{ {1, \ldots ,m} \right\}\backslash i} {\sum\limits_{l = 1}^{n_k } {\frac{{\partial f_{ijkl}^\textrm{pen} \left(\bm v \right)}}{{\partial \bm v_i }}} } }\\
\nabla _i F_\textrm{pro} \left(\bm v \right) &= \frac{\partial F_\textrm{pro} \left(\bm v_i \right)}{\partial\bm v_i} = \sum\limits_{j = 1}^{n_i } {\frac{\partial f_{ij}^{\textrm{pro}}\left( {\bm v_i } \right)}{\partial\bm v_i }}\\
\end{split}
\end{equation}

\begin{figure}[!htb]
	\centering
	\subfigure[ 
	1$^\textrm{st}$ iteration $F(\bm v) = 13741.2$]{
		\includegraphics[width=0.44\linewidth]{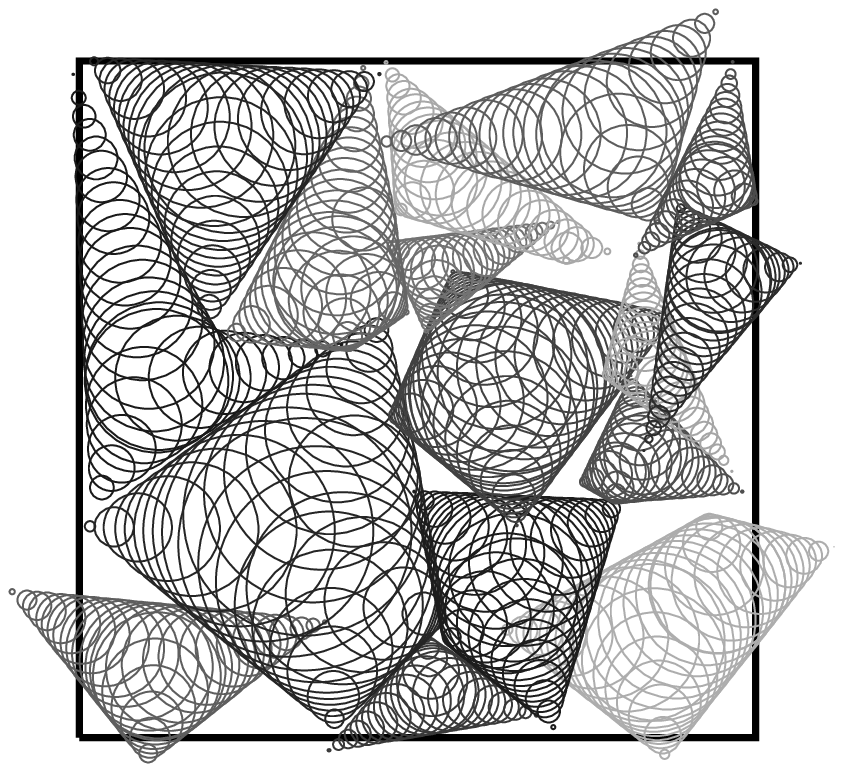}
		\label{fig:Evo_conv_algo_separa_ILSQN_a}
	}\quad
	\subfigure[2$^\textrm{nd}$ iteration $F(\bm v) = 6130.7$]{
		\includegraphics[width=0.44\linewidth]{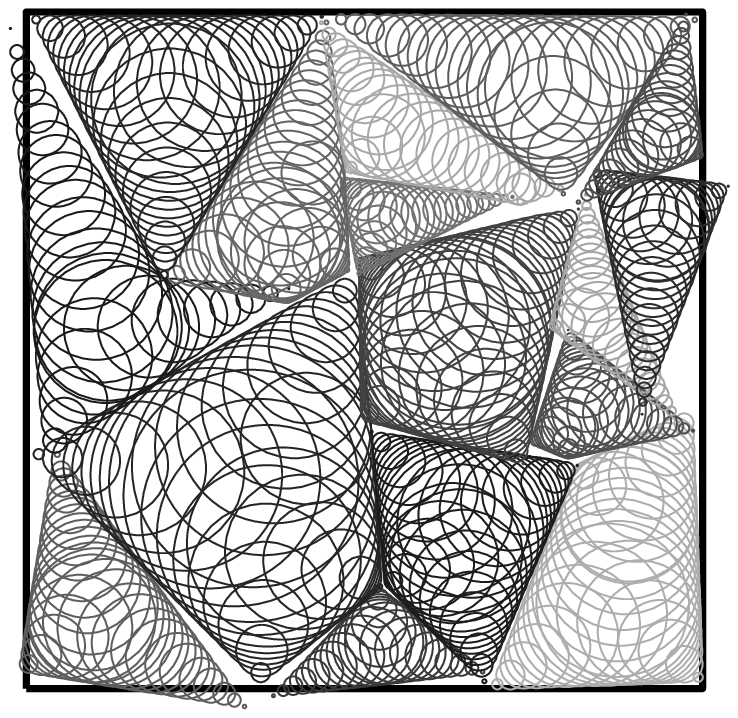}
		\label{fig:Evo_conv_algo_separa_ILSQN_b}
	}
    \\
	\subfigure[3$^\textrm{th}$ iteration $F(\bm v) = 904.1$]{
		\includegraphics[width=0.44\linewidth]{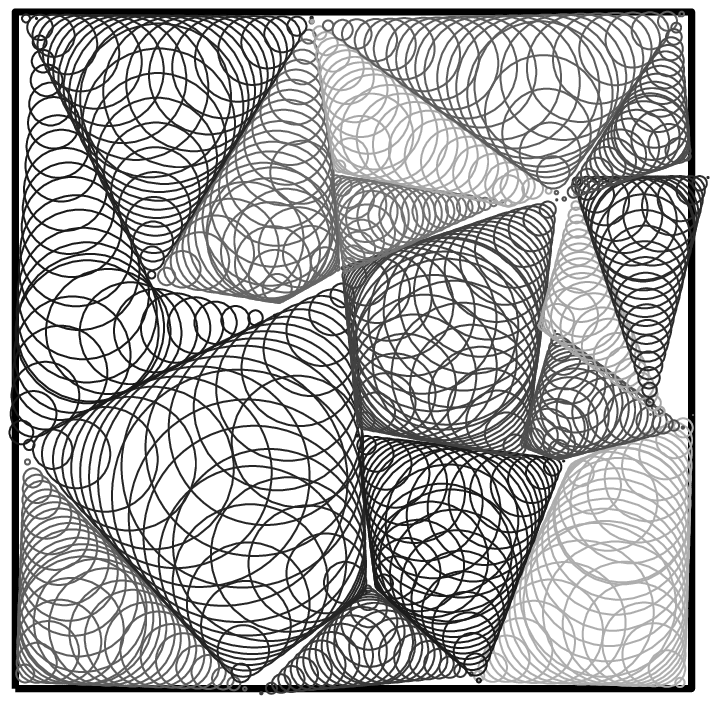}
		\label{fig:Evo_conv_algo_separa_ILSQN_c}
	}
	\quad
	\subfigure[4$^\textrm{th}$ iteration $F(\bm v) = 342.0$]{
		\includegraphics[width=0.44\linewidth]{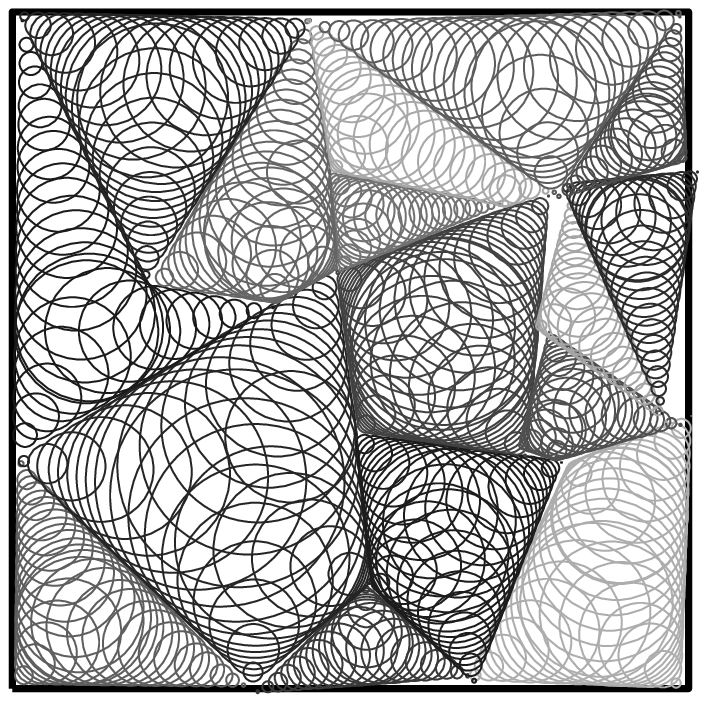}
		\label{fig:Evo_conv_algo_separa_ILSQN_d}
	}
    \\
	\subfigure[5$^\textrm{th}$ iteration $F(\bm v) = 122.4$]{
		\includegraphics[width=0.44\linewidth]{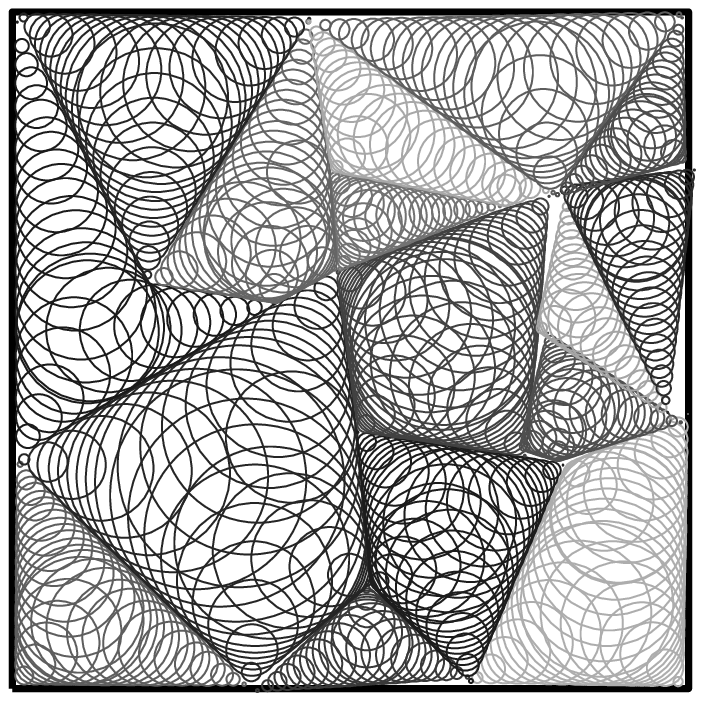}
		\label{fig:Evo_conv_algo_separa_ILSQN_e}
	}\quad
	\subfigure[6$^\textrm{th}$ iteration $F(\bm v) = 52.9$]{
		\includegraphics[width=0.44\linewidth]{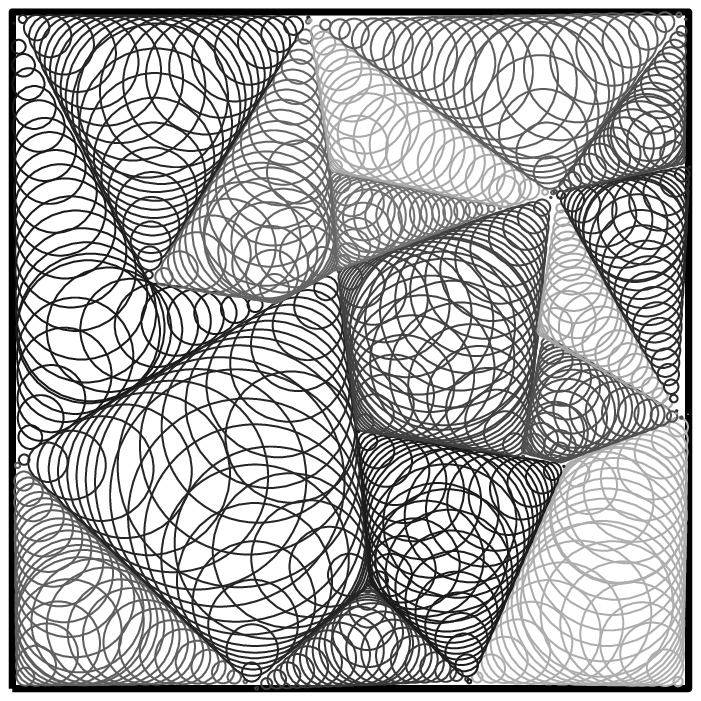}
		\label{fig:Evo_conv_algo_separa_ILSQN_f}
	}	
    \\
	\subfigure[20$^\textrm{th}$ iteration $F(\bm v) = 7\times10^{-3}$]{
		\includegraphics[width=0.44\linewidth]{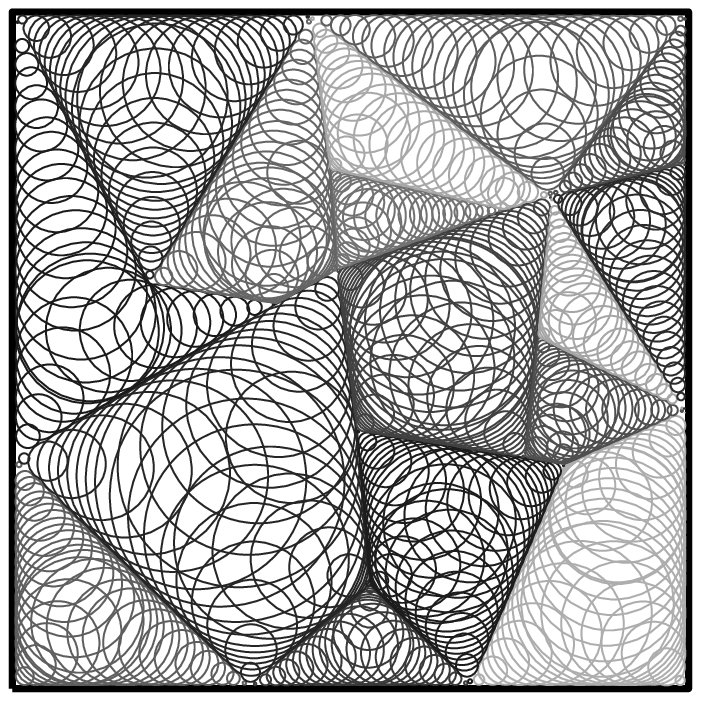}
		\label{fig:Evo_conv_algo_separa_ILSQN_k}
	}\quad
	\subfigure[30$^\textrm{th}$ iteration $F(\bm v) = 3\times10^{-4}$]{
		\includegraphics[width=0.44\linewidth]{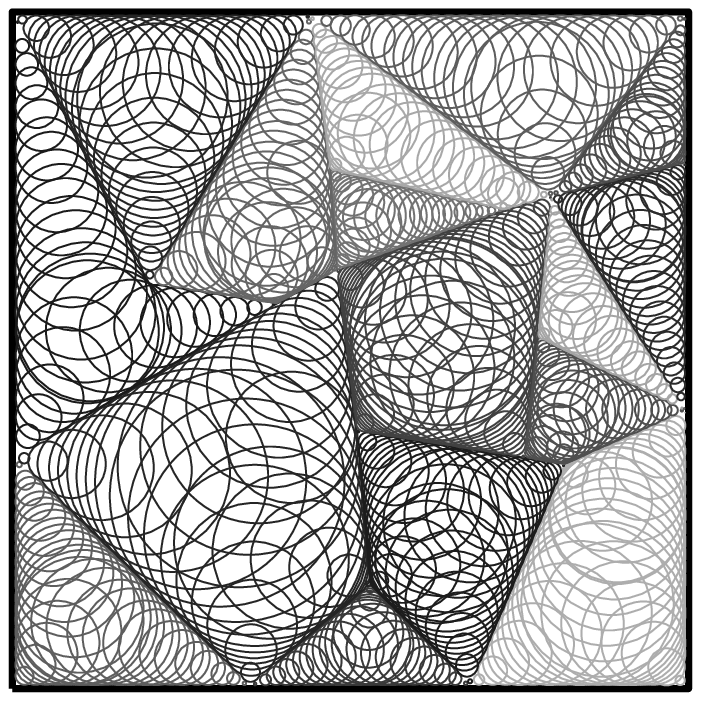}
		\label{fig:Evo_conv_algo_separa_ILSQN_l}
	}
	\caption{\MakeUppercase{Separation algorithm at work on instance \emph{Dighe~1} \cite{Digh_96_sol}. 16 components modelled with 515 circles. Each component has one rotation dof.}}
\label{fig:Evo_conv_algo_separa_ILSQN}
\end{figure}

For further details on the expressions of gradients, the reader can refer to Imamichi \etal \cite{Imam_08_des}.
The separation problem is solved using the BFGS quasi-Newton method \cite{LiuN_89_lim}.
This technique is iterative and moves all components at the same time,
unlike different separation heuristics that move one component at a time \cite{Egeb_07_fas}.
Figure \ref{fig:Evo_conv_algo_separa_ILSQN} illustrates how the separation algorithm works on a solution proposed by the global optimizer.
The order of magnitude of the function $F$ depends on the circle conversion.
However, when placement constraints are satisfied, the function $F$ should always be zero.

The separation algorithm allows one to find a feasible solution from a solution that does not satisfy placement constraints.
The generation of solutions is the role of the global optimizer.

\subsection{Global optimizer}
The global optimizer is in charge of exploring efficiently the search space to propose promising solutions.
The genetic algorithm (GA) Omni-Optimizer \cite{DebK_08_omn} is used. Its objective is to improve the current population at each new generation.
Based on NSGA-2 \cite{DebK_00_fas}, this generational GA is designed to handle single and multi-objective problems.
This algorithm was originally chosen for its capacities to find multi-modal solutions. Three variation operators are used to generate new solutions:
\begin{enumerate}
	\item The crossover operator takes two solutions and generates two offspring solutions by crossing over all parent variables under a certain probability. The standard crossover scheme \emph{SBX} is used for real variables \cite{DebK_95_sim} and the two point crossover is used for discrete variables. The crossover operator is used with extreme parsimony: indeed when the compacity of the problem increases, crossing over two solutions do not produce promising solutions (\ie offspring solutions present too much overlaps when compared with their parent solutions), even when using the restricted selection operator. Introduced by Deb \etal \cite{DebK_08_omn}, the restricted selection operator consists in selecting two parents which genotypic distance is minimum.
	\item Mutations generate new solutions from one parent. The polynomial mutation operator is used for real variables, and a bit wise mutation is used for discrete variables.
	\item A swap operator that exchanges the position of two components has been added. This classical operator in placement problems allows one to generate new solutions from one parent solution.
\end{enumerate}

\subsection{The proposed algorithm}
\begin{figure}[!ht]
	\begin{center}
		\begin{psfrags}
			\psfrag{Pi}[c][c]{$P_i$}
			\psfrag{i=0}[c][c]{$i=0$}			
			\psfrag{i=i+1}[c][c]{$i=i+1$}			
			\psfrag{P0}[c][c]{$P_0$}	
			\psfrag{Algorithme genetique}[c][c]{}
            \includegraphics[width=0.95\linewidth]{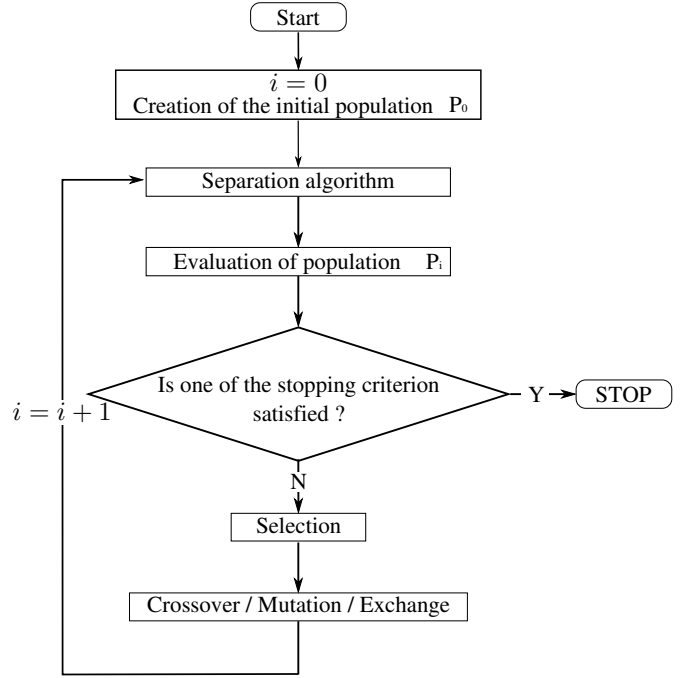}
		\end{psfrags}
	\end{center}
    \caption{Genetic algorithm proposed to solve placement problems.}
	\label{fig:Principes_algo_gene}
\end{figure}

The proposed algorithm is presented figure \ref{fig:Principes_algo_gene}.
The structure of the algorithm is very close to a generational genetic algorithm.
The separation algorithm is nested in the genetic algorithm, and modifies component position so that the solution proposed respect placement constraints.
The initial population can be generated randomly, or the designer can provide a set of solutions from its choice or also solutions obtained from a placement heuristic such as bottom-left placement technique.

Before evaluating a solution, the algorithm checks if placement constraints are satisfied.
If so, the different objectives of the solution are evaluated and the algorithm moves to the next solution.
Otherwise, the separation algorithm is run and modifies the solution so that placement constraints are respected.
The solution is then evaluated.
At the end of the local optimization performed by the separation algorithm,
the value of function $F$ returned is used as a constraint violation indicator for the genetic algorithm.
This indicator will then be taken into account in the genetic operations.
As a consequence, a solution that does not respect placement constraints will not be selected in the constrained binary tournament when compared to a feasible solution.


\subsection{Characteristics of the method}
This subsection sums up the different characteristics and possibilities of the proposed method:
\begin{enumerate}
	\item Components can be restricted to lay in an area defined by the designer. It consists in reducing the inward offset domains of a component.
	\item Alignement constraints can be included in the modelling and in the separation algorithm, therefore automatically being satisfied.
	\item The designer can easily interact with solutions, and can propose solutions even if these solutions do not respect placement constraints.
	\item Distance queries between components can easily be satisfied. For example, a minimum distance $d$ between two components $i$ and $k$ can be obtained by virtually growing radius of each circle of components $i$ and $k$ of $d/2$ when considering their possible overlap.
	\item In the case, components are all rectangular in 2D and parallelepipedic in 3D, the
	separation algorithm can be adapted to handle these special cases.
	The objective of the separation algorithm may be to reduce the sum of overlaping areas/volumes,
	for which an analytical function and gradient can easily be found.
	\item The separation algorithm may be optional for problems with a low compacity. In this case, the method is equivalent to use a simple genetic algorithm to solve a placement problem, as Yi \etal \cite{YiFa_07_veh} already did.
\end{enumerate}

\section{Simulation results}
\subsection{Data of the problem}
Let's consider the problem of loading a car trunk, in which a set of $11$ components should be placed (Figures \ref{fig:Trunk}a \& \ref{fig:Trunk}b). Two objectives are taken into account: the first one consists in minimizing the euclidean distance between the geometric center of gravity of the trunk and the center of gravity of the assembly. The second objective consists in minimizing the inertia moment of the assembly around the $y$ axis of the trunk (Figure \ref{fig:Trunk}a). The first objective gives information on the static behavior of the trunk, whereas the second is more about the dynamic behavior.

Component properties are given in table \ref{tab:comp_pro}.
The grayscale of components will be used to code their density: the darker a component, the larger its density.
The global compacity of this problem is evaluated to 79.5\%.
Orientations of components are chosen to be discrete.
For each component, the set of possible orientations is $\left\{ {0^\circ,90^\circ,180^\circ,270^\circ} \right\}$, except for symetrical components which orientation is chosen among the set $\left\{ {0^\circ,90^\circ} \right\}$.

The problem has $22$ real variables coding component positions and $11$ discrete variables coding component orientations. The separation algorithm only translates components to legalize solutions proposed by the global optimizer. The formulation of the problem can be formulated as follows:
\begin{equation}
	\mathscr P \left\{
	\begin{array}{l}
		\min f_1\left(\bm v\right)=\sqrt {d_x^2  + d_y^2 }  \\	
		\min f_2\left(\bm v\right) = \sum\limits_{i = 1}^m {\rho _i I_{i/yy} }\\ 
		\textrm{s.t. placement constraints are satisfied}
	\end{array}
	\right.
\end{equation}
where
\begin{equation}
	\begin{array}{l}
		d_x  = \frac{1}{{\sum\limits_{i = 1}^m {\rho _i A_i } }}\sum\limits_{i = 1}^m {\rho _i A_i \left( {x_i  - x_G } \right)}\\
		d_y  = \frac{1}{{\sum\limits_{i = 1}^m {\rho _i A_i } }}\sum\limits_{i = 1}^m {\rho _i A_i \left( {y_i  - y_G } \right)}\\
		I_{i/yy}  = \iint\limits_{(x,y) \in P_i } {\left( {x - x_G } \right)}^2 dxdy\\
	\end{array}\\
\end{equation}
with $(x_G,y_G)$ the cordinates of the geometric center of gravity.

Parameters of the genetic algorithm are given table \ref{tab:param_GA}.
The probability of crossover of real and discrete variables are very low:
in this example, the global optimizer looks like more an evolution strategy than a genetic algorithm.
These parameters are set to such low values because the high compacity of problem does not permit to generate promising solutions with crossovers.
For high compacity problems, the priority is given to mutation and swap operators.

\begin{figure*}
	\centering
	\subfigure[Trunk geometry]{
		\includegraphics[width=0.27\linewidth]{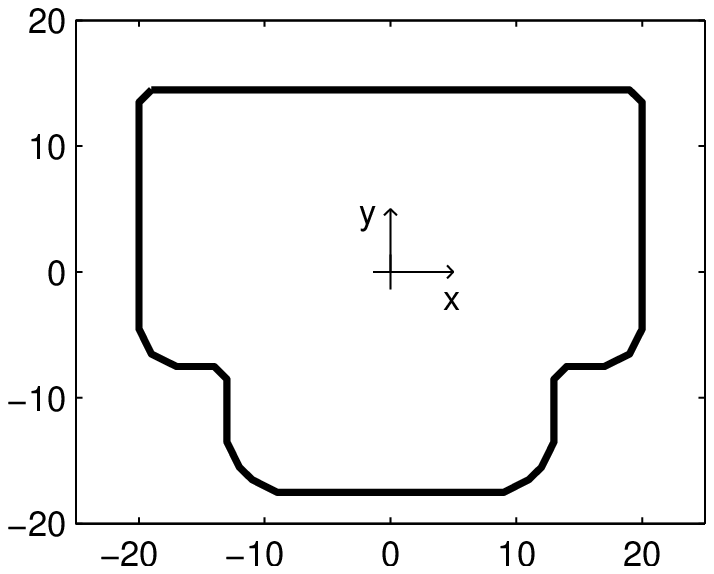}
		\label{fig:Trunk_a}
	}
	\quad
	\subfigure[Components]{
		\includegraphics[width=0.27\linewidth]{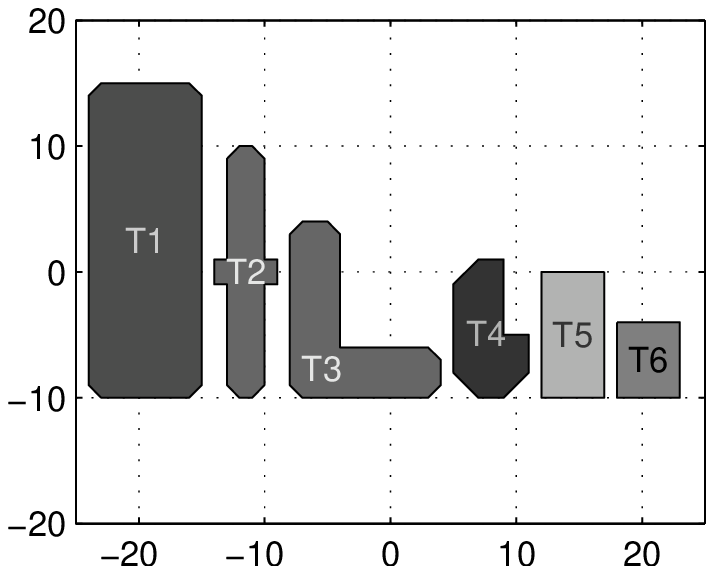}
		\label{fig:Trunk_b}
	}
    \if@twocolumn
        \caption{\MakeUppercase{presentation of the container and the different components of the example.}}
        \label{fig:Trunk}
    \else
        \caption{Presentation of the container and the different components of the example.}
        \label{fig:Trunk}
    \fi
\end{figure*}

\begin{table}[t]
    \if@twocolumn
        \caption{COMPONENT PROPERTIES}
        \label{tab:comp_pro}
    \else
        \caption{Component properties}
        \label{tab:comp_pro}
    \fi
	\begin{center}
		\begin{tabular}{|l|c|c|c|c|c|c|}
		\hline
		Type of components   &T1 & T2 & T3 & T4 & T5 & T6\\  \hline
		Number of components & 2 & 2 & 1 & 1 & 2 & 3\\ \hline
		Density of components&0.7 & 0.6 & 0.6 & 0.8 & 0.3 & 0.5\\ \hline
		\end{tabular}
	\end{center}
\end{table}

\begin{table}[t]
\if@twocolumn
    \caption{\MakeUppercase{Parameters of the GA}}
    \label{tab:param_GA}
\else
    \caption{Parameters of the GA}
    \label{tab:param_GA}
\fi
\begin{center}
	\begin{tabular}{|l|c|}
		\hline
		Number of generations &  100 \\ \hline
		Number of individuals  & 100 \\ \hline
		Crossover probability of real variables & 0.05  \\ \hline
		Mutation probability of real variables & 0.4  \\ \hline
		Crossover probability of binary/discrete variables & 0  \\ \hline
		Mutation probability of binary/discrete variables  &0.3  \\ \hline
		Distribution index $\eta_c$ for real variable crossover&5  \\ \hline
		Distribution index $\eta_m$ for real variable mutation&5  \\ \hline
		Swapping probability of two components &0.05  \\ \hline
		Relative coefficient used for the $\varepsilon$-domination  &0.001  \\ \hline
		Use of phenotypic and genotypic distance  &Yes  \\ \hline
		Use of restricted selection operator  &Yes  \\ \hline
	\end{tabular}
\end{center}
\end{table}

\subsection{Results and analysis}
Figure \ref{fig:solutions} presents a set of efficient solutions extracted from the Pareto front of figure \ref{fig:pareto_front}.
Subfigures are sorted in the increasing order of objective $1$ and decreasing order for objective $2$.
On each solution presented, every component is in contact with another one, and one can observe that there is very little space to move components.
One can also observe that the darkest components, \ie the components with the largest density, are located close to the center of the trunk or along the $y$-axis:
these components have a larger impact on the objective functions and by placing them close to the geometric center of gravity of the trunk, one optimizes both objective functions.
Once the Pareto front is found, a multi-criteria analysis can be performed to choose one solution.
\begin{figure*}
	\centering
	\subfigure[{$f_1 = 0.002, f_2 = 48204.0$}]
	{
		\includegraphics[width=0.27\linewidth]{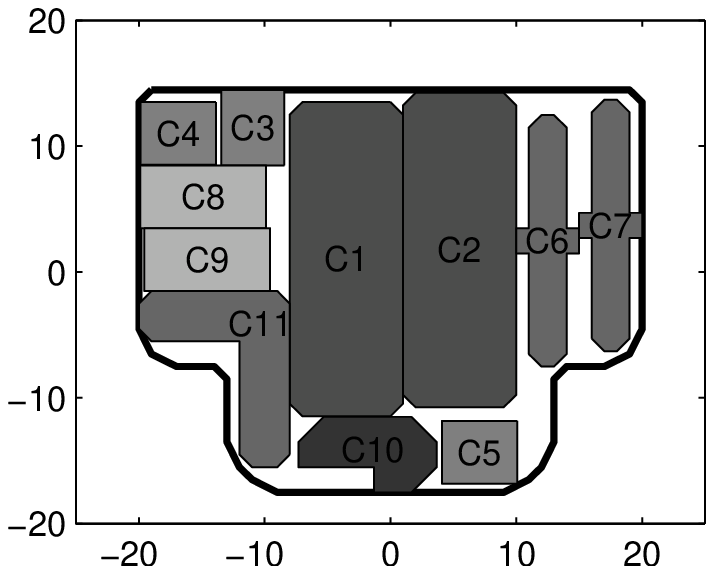}
		\label{fig:Trunk_sol1}
	}
	\quad
	\subfigure[{$f_1 = 0.010, f_2 = 45274.5$}]
	{
		\includegraphics[width=0.27\linewidth]{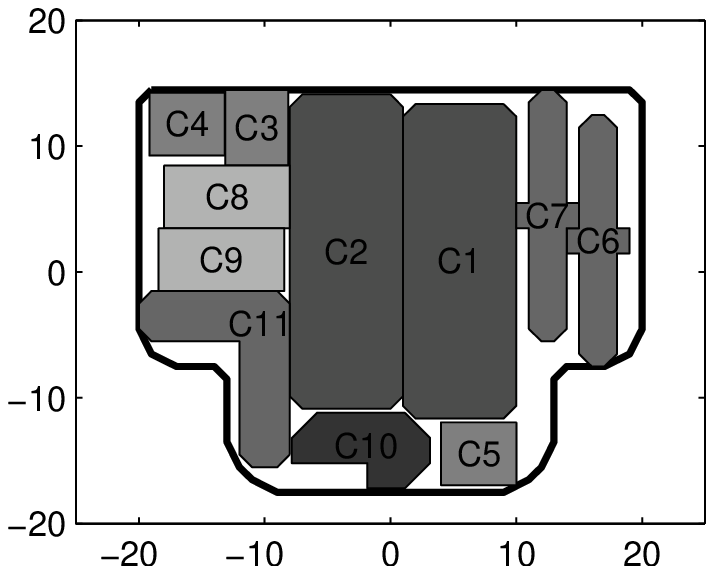}
		\label{fig:Trunk_sol2}
	}
	\quad
	\subfigure[{$f_1 = 0.1052, f_2 = 47277.3$}]
	{
		\includegraphics[width=0.27\linewidth]{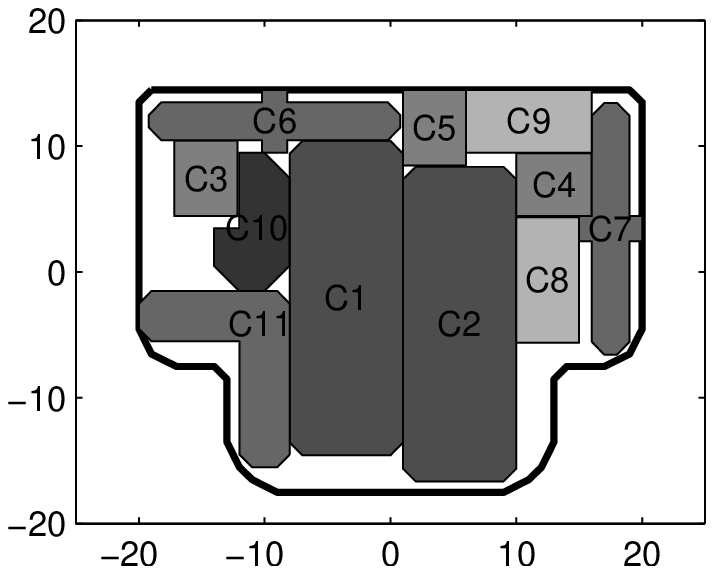}
		\label{fig:Trunk_sol3}
	}
	\\
	\subfigure[{$f_1 = 0.506, f_2 = 44899.6$}]
	{
		\includegraphics[width=0.27\linewidth]{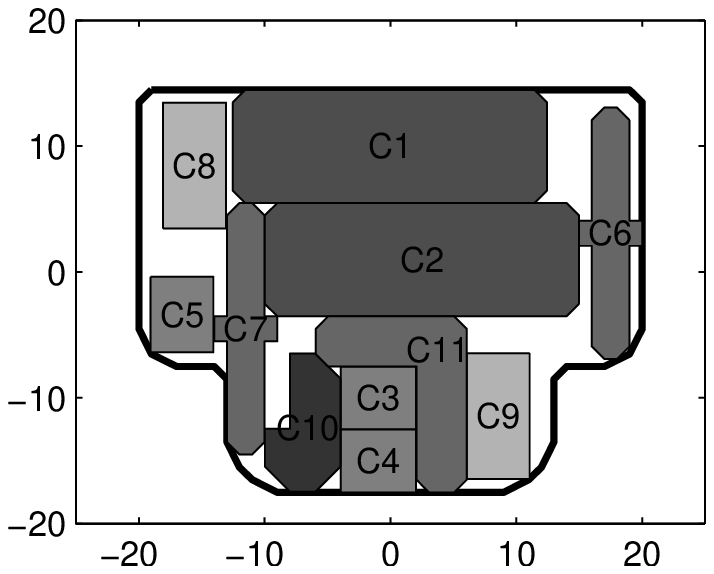}
		\label{fig:Trunk_sol4}
	}
	\quad
	\subfigure[{$f_1 = 0.600, f_2 = 43973.2$}]
	{
		\includegraphics[width=0.27\linewidth]{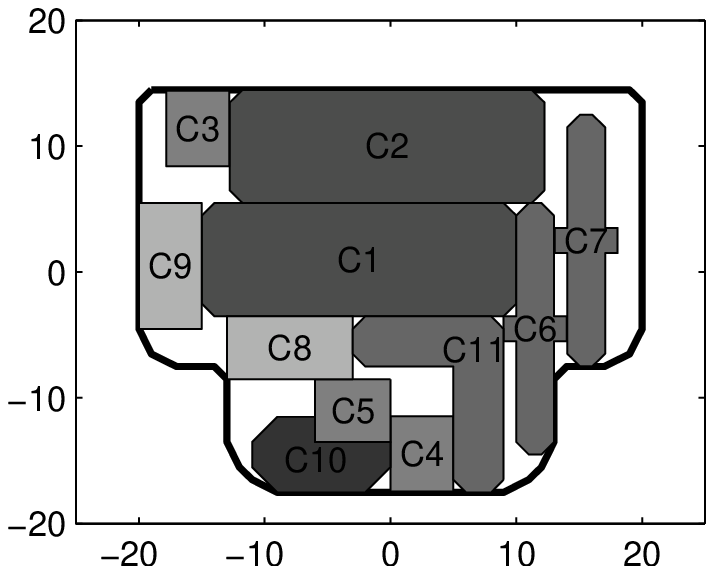}
		\label{fig:Trunk_sol5}
	}
	\quad
	\subfigure[{$f_1 = 0.900, f_2 = 41239.5$}]
	{
		\includegraphics[width=0.27\linewidth]{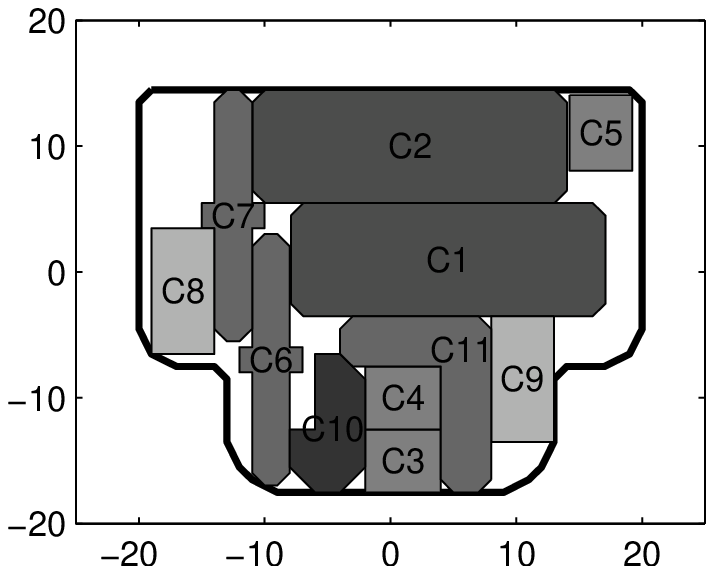}
		\label{fig:Trunk_sol6}
	}
    \if@twocolumn
        \caption{\MakeUppercase{Example of solutions extracted from the Pareto front.}}
        \label{fig:solutions}
    \else
        \caption{Example of solutions extracted from the Pareto front.}
        \label{fig:solutions}
    \fi
\end{figure*}

Figure \ref{fig:pareto_front} presents the different trade-off surfaces obtained for the different simulations.
One simulation corresponds to one execution of the genetic algorithm. Each simulation evaluates $10,000$ solutions in $15$ minutes.
The problem is solved with $13$ different initial random populations. Each point corresponds to a feasible solution, meaning that the corresponding solution satisfies placement constraints.
Identical symbols represent the trade-off surface obtained from one simulation, \ie the Pareto set obtained.
The first conclusion that can be drawn from this figure is that there are large disparities between results.
Two different initial populations do not converge towards the same trade-off surface. If one has a look at the solutions proposed at the end of a simulation, one can see that the layouts associated with these solutions are very close to each other, meaning that diversity of the population in the variable space and in the objective space has been lost.
The loss of diversisty can be explained by the fact that the first feasible solutions found generate a selection pressure on the other unfeasible solutions.
As a consequence, unfeasible but promising solutions are not selected for next generations, making them disappear. After a certain number of generations, all solutions begin to look like each other and the convergence stalls, the genetic operators do not succeed in proposing new efficient solutions. Deb \etal \cite{DebK_08_omn} have introduced the concept of $\varepsilon$-domination to preserve diversity in the population. This relaxation on the dominance relation allows inferior solutions to be kept in the population while preserving diversity. This operator could be extended to constraints in order to preserve diversity of population for problems with high compacity. It may also be interesting to regenerate a part of the population when one detects that diversity is being lost. Tiwari \etal \cite{Tiwa_08_fas} use this feature inside their steady-state genetic algorithm.

\begin{figure*}
	\centering
	\psfragscanon
	\psfrag{F}[c][c]{$f$}
	\includegraphics[width=0.6\linewidth]{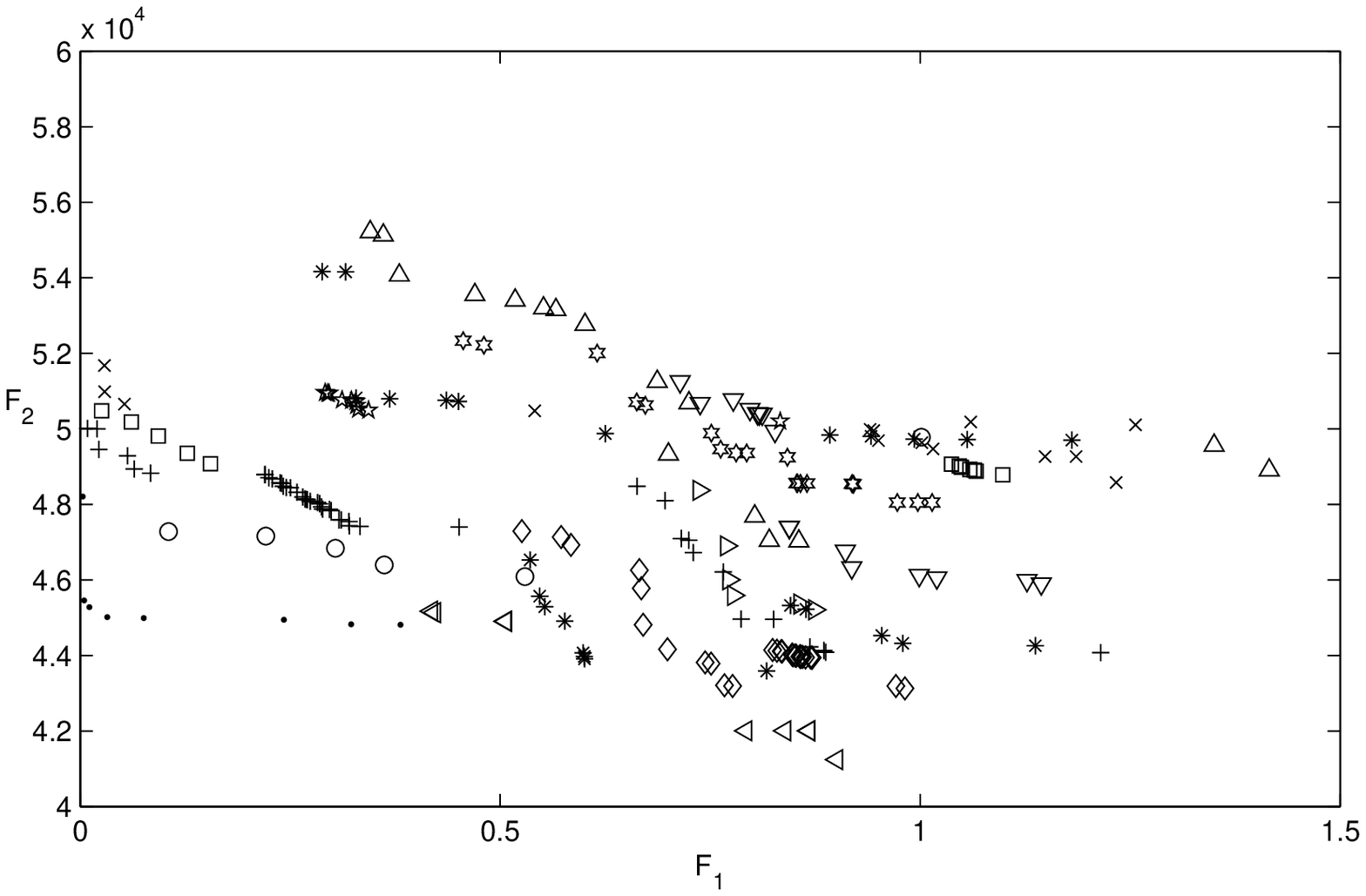}
    \if@twocolumn
        \caption{\MakeUppercase{Presentation of different Pareto fronts obtained for different seeds.}}
        \label{fig:pareto_front}
    \else
        \caption{Presentation of different Pareto fronts obtained for different seeds.}
        \label{fig:pareto_front}
    \fi
\end{figure*}

Numerical results tend to prove that the higher the compacity of the problem is, the harder it is to find the set of efficient solutions, and once a feasible solution has been found it is difficult to change completely the topology of the solution. If the same problem is solved without one component of type 1, the compacity falls to 59.6\%, and the trade-off surfaces obtained are identical whatever the initial population. When compacity of the problem grows, the problem transforms itself into a C\&P problem, where the main objective is to find a feasible solution.

\section{3D separation algorithm}
The proposed method can be adapted to 3D, the difficulty being in the adaptation of the separation algorithm. The genetic algorithm can easily be adapted to the 3D version of the problem.
Imamichi \etal \cite{Imam_08_des} has already adapted his separation algorithm to 3D, however only with simple components and enclosure.
To adapt the separation algorithm, the first task consists in converting polyhedra in sets of spheres.
Hubbard \cite{Hubb_96_app} and Bradshaw \etal \cite{Brad_00_ext} have already solved this problem by approximating polyhedra with hierarchies of spheres based on component medial-axis surfaces (skeletal representations of components).
Non-overlap constraints can be checked by simple distance computations between sphere centers.
Non-protrusion constraints require to compute the minimum distance between sphere centers and the boundary of the enclosure.
The challenge of such an algorithm is to find the correct trade-off between the number of spheres required to represent components and the computational complexity of the collision detection.
Preliminary studies reveal that only decompositions based on medial axis provide interesting results.
Other methods such as sphere-trees based on octree decomposition are prone to premature convergence during the separation optimization.

\section{Conclusion}
This paper has introduced a 2D multi-objective placement method for complex geometry components.
The proposed relaxed placement technique is based on the hybridation of a genetic algorithm and a separation algorithm, and allows one to solve placement problems with several types of placement constraints.
Applications of the proposed method can be found in engineering domains, where placement problems have no particular specificities and can not be treated with classical methods.

Results show that high quality solutions can be obtained with appropriate parameters for the genetic algorithms.
Immediate work includes a study of the influence of the initial population and a statistical study of the influence of the parameters of the genetic algorithm.
Several tests need to be performed to see if diversity can be maintained inside the population of solutions.
Trade-off surfaces obtained for different simulations must be numerically evaluated with multi-objective indicators (\eg delineation, distance, diversity and hypervolume) to quantify the differences between them.

Several extensions can be included in this work: the collision detection process can be speeded up by using a hierarchical description of components, articulated components could be taken into account with the circle decomposition of components.
Future works also include comparison of different modelling techniques for high compacity problems: indeed placement problems with a large compacity are close to packing problems, for which fast and efficient technique have been proposed \cite{Tiwa_08_fas}.

\bibliographystyle{asmems4}

\begin{thebibliography}{10}

\bibitem{Alad_07_obj1}
Aladahalli, C., Cagan, J., and Shimada, K., 2007.
\newblock ``Objective function effect based pattern search - theoretical
  framework inspired by 3{D} component layout''.
\newblock {\em {J}ournal of {M}echanical {D}esign, {\bf 129}}(3), Mar.,
  pp.~243--254.

\bibitem{Wasc_06_imp}
W\"ascher, G., Hau{\ss}ner, H., and Schumann, H., 2007.
\newblock ``An improved typology of cutting and packing problems''.
\newblock {\em {E}uropean {J}ournal of {O}perational {R}esearch, {\bf 183}},
  Dec., pp.~1109--1130.

\bibitem{Dyck_90_typ}
Dyckhoff, H., 1990.
\newblock ``A typology of cutting and packing problems''.
\newblock {\em {E}uropean {J}ournal of {O}perational {R}esearch, {\bf 44}}(2),
  Jan., pp.~145--159.

\bibitem{Kell_04_kna}
Kellerer, H., Pferschy, U., and Pisinger, D., 2004.
\newblock {\em Knapsack Problems}.
\newblock Springer, Berlin, Germany.

\bibitem{Carl_07_new}
Carlier, J., Clautiaux, F., and Moukrim, A., 2007.
\newblock ``New reduction procedures and lower bounds for the two-dimensional
  bin packing problem with fixed orientation''.
\newblock {\em {C}omputers \& {O}perations {R}esearch, {\bf 34}}(8), Aug.,
  pp.~2223--2250.

\bibitem{Egeb_07_fas}
Egeblad, J., Nielsen, B.~K., and Odgaard, A., 2007.
\newblock ``Fast neighborhood search for two- and three-dimensional nesting
  problems''.
\newblock {\em {E}uropean {J}ournal of {O}perational {R}esearch, {\bf 127}}(3),
  Dec., pp.~1249--1266.

\bibitem{Gome_06_sol}
Gomes, M.~A., and Oliveira, J.~F., 2006.
\newblock ``Solving irregular strip packing problems by hybridising simulated
  annealing and linear programming''.
\newblock {\em {E}uropean {J}ournal of {O}perational {R}esearch, {\bf 171}}(3),
  June, pp.~811--829.

\bibitem{Pisi_02_heu}
Pisinger, D., 2002.
\newblock ``Heuristics for the container loading problem''.
\newblock {\em {E}uropean {J}ournal of {O}perational {R}esearch, {\bf 141}},
  pp.~382--392.

\bibitem{Drir_07_fac}
Drira, A., Pierreval, H., and Hajri-Gabouj, S., 2007.
\newblock ``Facility layout problems: A survey''.
\newblock {\em Annual Reviews in Control, {\bf 31}}(2), pp.~255--267.

\bibitem{Egeb_03_pla}
Egeblad, J., 2003.
\newblock ``Placement techniques for {VLSI} layout using sequence-pair
  legalization''.
\newblock Master's thesis, Department of Computer Science, University of
  Copenhagen, July.

\bibitem{Zhan_08_lay}
Zhang, B., Teng, H.-F., and Shi, Y.-J., 2008.
\newblock ``Layout optimization of satellite module using soft computing
  techniques''.
\newblock {\em {A}pplied {S}oft {C}omputing, {\bf 8}}(1), pp.~507--521.

\bibitem{Caga_02_sur}
Cagan, J., Shimada, K., and Yin, S., 2002.
\newblock ``A survey of computational approaches to three-dimensional layout
  problems.''.
\newblock {\em Computer-Aided Design, {\bf 34}}(8), pp.~597--611.

\bibitem{Grig_04_gab}
Grignon, P.~M., and Fadel, G.~M., 2004.
\newblock ``A {GA} based configuration design optimization method''.
\newblock {\em {J}ournal of {M}echanical {D}esign, {\bf 126}}(1), Jan.,
  pp.~6--15.

\bibitem{Alad_07_obj2}
Aladahalli, C., Cagan, J., and Shimada, K., 2007.
\newblock ``Objective function effect based pattern search - an implementation
  for 3{D} component layout''.
\newblock {\em {J}ournal of {M}echanical {D}esign, {\bf 129}}(3), Mar.,
  pp.~255--265.

\bibitem{YiFa_07_veh}
Miao, Y., Fadel, G.~M., and Gantovnik, V.~B., 2008.
\newblock ``Vehicle configuration design with a packing genetic algorithm''.
\newblock {\em International Journal of Heavy Vehicle Systems, {\bf 15}}, 12,
  pp.~433--448(16).

\bibitem{Tiwa_08_fas}
Tiwari, S., Fadel, G., and Fenyes, P., 2008.
\newblock ``A fast and efficient three dimensional compact packing algorithm
  for free-form objects''.
\newblock In Proceedings of ASME International Design Engineering \& Computers
  and Information in Engineering Conference (ASME DETC/CIE 2008).
\newblock DETC2008-50097.

\bibitem{Dong_06_veh}
Dong, H., Fadel, G.~M., and Blouin, V.~Y., 2006.
\newblock ``Vehicle component layout with shape morphing - an initial study''.
\newblock In ASME International Design Engineering Technical Conferences \&
  Computers and Information in Engineering Conference.

\bibitem{Tiwa_06_sur}
Tiwari, S., Fadel, G., and Gantovnik, V., 2006.
\newblock ``A survey of various encoding schemes and associated placement
  algorithms applied to packing and layout problems''.
\newblock In Proceedings of IDETC/CIE. ASME 2006 International Design
  Engineering Technical Conferences \& Computers and Information in Engineering
  Conference.

\bibitem{Benn_08_com}
Bennell, J.~A., and Song, X., 2008.
\newblock ``A comprehensive and robust procedure for obtaining the nofit
  polygon using minkowski sums''.
\newblock {\em {C}omputers \& {O}perations {R}esearch, {\bf 35}}(1),
  pp.~267--281.

\bibitem{Imam_07_ite}
Imamichi, T., Yagiura, M., and Nagamochi, H., 2007.
\newblock ``An iterated local search algorithm based on nonlinear programming
  for the irregular strip packing problem''.
\newblock In Proceedings of the Third International Symposium on Scheduling,
  Tokyo, Japan, no.~9, pp.~132--137.

\bibitem{Imam_07_mul}
Imamichi, T., and Nagamochi, H., 2007.
\newblock ``A multi-sphere scheme for 2{D} and 3{D} packing problems''.
\newblock In SLS, pp.~207--211.

\bibitem{Imam_08_rem}
Imamichi, T., Arahori, Y., Gim, J., Hong, S.-H., and Nagamochi, H., 2008.
\newblock Removing overlaps in label layouts using multi-sphere scheme.
\newblock Tech. Rep. 2008-006, Department of Applied Mathematics and Physics,
  Kyoto University and School of Information Technologies, University of
  Sydney.

\bibitem{Imam_08_des}
Imamichi, T., and Nagamochi, H., 2008.
\newblock ``Designing algorithms with multi-sphere scheme''.
\newblock In ICKS '08: Proceedings of the International Conference on
  Informatics Education and Research for Knowledge-Circulating Society (icks
  2008), IEEE Computer Society, pp.~125--130.

\bibitem{Agar_00_pen}
Agarwal, P.~K., Guibas, L.~J., Har-Peled, S., Rabinovitch, A., and Sharir, M.,
  2000.
\newblock ``Penetration depth of two convex polytopes in {3D}''.
\newblock {\em Nordic Journal of Computing, {\bf 7}}(3), pp.~227--240.

\bibitem{LiuN_89_lim}
Liu, D.~C., and Nocedal, J., 1989.
\newblock ``{On the limited memory BFGS method for large scale optimization}''.
\newblock {\em Math. Program., {\bf 45}}(3), pp.~503--528.

\bibitem{DIGH_96_SOL}
Dighe, R., and Jakiela, M.~J., 1996.
\newblock ``Solving pattern nesting problems with genetic algorithms employing
  task decomposition and contact detection''.
\newblock {\em Evolutionary {C}omputation, {\bf 3}}(3), pp.~239--266.

\bibitem{DebK_08_omn}
Deb, K., and Tiwari, S., 2008.
\newblock ``{O}mni-{O}ptimizer: A generic {E}volutionary {A}lgorithm for single
  and multi-objective optimization''.
\newblock {\em {E}uropean {J}ournal of {O}perational {R}esearch, {\bf 185}},
  Mar., pp.~1062--1087.

\bibitem{DebK_00_fas}
Deb, K., Agrawal, S., Pratab, A., and Meyarivan, T., 2000.
\newblock ``A fast elitist {N}on-dominated {S}orting {G}enetic {A}lgorithm for
  multi-objective optimization: {NSGA-{II}}''.
\newblock In Proceedings of the Parallel Problem Solving from Nature {VI}
  Conference, Springer. Lecture Notes in Computer Science No. 1917,
  pp.~849--858.

\bibitem{DebK_95_sim}
Deb, K., and Agrawal, R.~B., 1995.
\newblock ``Simulated binary crossover for continuous search space''.
\newblock {\em Complex Systems, {\bf 9}}, pp.~115--148.

\bibitem{Hubb_96_app}
Hubbard, P.~M., 1996.
\newblock ``Approximating polyhedra with spheres for time-critical collision
  detection''.
\newblock {\em ACM Transactions on Graphics, {\bf 15}}(3), pp.~179--210.

\bibitem{Brad_00_ext}
Bradshaw, G., and O'sullivan, C., 2000.
\newblock ``Extracting geometric models from medieval moulding profiles for
  case-based reasoning.''.
\newblock In KES, R.~J. Howlett and L.~C. Jain, eds., IEEE, pp.~531--536.

\end{thebibliography}

\end{document}